\documentclass[twocolumn,superscriptaddress,amsmath,amssymb]{revtex4}
\usepackage{graphicx}
\usepackage{dcolumn}
\usepackage{bm}
\usepackage{ulem}
\newcommand{\nc}{\newcommand}
\nc{\be}{\begin{eqnarray}}
\nc{\ee}{\end{eqnarray}}
\nc{\bea}{\begin{eqnarray}}
\nc{\eea}{\end{eqnarray}}
\nc{\bean}{\begin{eqnarray*}}
\nc{\eean}{\end{eqnarray*}}

\begin{document}

\title{
Emergent photovoltage on SmB$_6$ surface upon bulk-gap evolution 
revealed by pump-and-probe photoemission spectroscopy
}

\author{Y.~Ishida}
\affiliation{ISSP, University of Tokyo, Kashiwa-no-ha, Kashiwa, 
Chiba 277-8561, Japan}

\author{T.~Otsu}
\affiliation{ISSP, University of Tokyo, Kashiwa-no-ha, Kashiwa, 
Chiba 277-8561, Japan}

\author{T.~Shimada}
\affiliation{ISSP, University of Tokyo, Kashiwa-no-ha, Kashiwa, 
Chiba 277-8561, Japan}

\author{M.~Okawa}
\affiliation{ISSP, University of Tokyo, Kashiwa-no-ha, Kashiwa, 
Chiba 277-8561, Japan}

\author{Y.~Kobayashi}
\affiliation{ISSP, University of Tokyo, Kashiwa-no-ha, Kashiwa, 
Chiba 277-8561, Japan}

\author{F.~Iga}
\affiliation{College of Science, Ibaraki University, Mito, Ibaraki 310-8512, Japan}
\affiliation{Department of Quantum Matter and Institute for Advanced Materials Research,
Hiroshima University, Higashi-hiroshima, Hiroshima 739-8530, Japan}

\author{T.~Takabatake}
\affiliation{Department of Quantum Matter and Institute for Advanced Materials Research,
Hiroshima University, Higashi-hiroshima, Hiroshima 739-8530, Japan}

\author{S.~Shin}
\affiliation{ISSP, University of Tokyo, Kashiwa-no-ha, Kashiwa, 
Chiba 277-8561, Japan}

\date{\today}

\begin{abstract}
Recent studies suggest that an exemplary Kondo insulator SmB$_6$ belongs to 
a new class of topological insulators (TIs), in which  
non-trivial spin-polarized metallic states emerge on surface upon the formation of 
Kondo hybridization gap in the bulk. 
Remarkably, the bulk resistivity reaches more than 20 $\Omega$\,cm at 4 K, 
making SmB$_6$ a candidate for a so-called bulk-insulating TI. 
We here investigate optical-pulse responses of SmB$_6$ by pump-and-probe 
photoemission spectroscopy. Surface photovoltage effect is observed below $\sim$90 K. 
This indicates that an optically-active band bending region develops 
beneath the novel metallic surface upon the bulk-gap evolution. 
The photovoltaic effect persists for $>$200 $\mu$s, 
which is long enough to be detected by electronics devices, 
and could be utilized for optical gating of the novel metallic surface. 
\end{abstract}

\maketitle 
Topological insulators (TIs) are promising spin-electronic device materials 
because they exhibit spin-polarized metallic states on surface \cite{Hasan_Rev}. 
The surface of TI is topologically constrained to become 
a two-dimensional massless Dirac electron system 
that shows novel properties such as high mobility and suppression of backscattering.  
Extensive research is being pursued to extract the surface-dominated conduction 
by realizing a so-called intrinsic TI, 
in which the bulk shows insulating behavior 
\cite{Bi2Se2Te_PRB10, BSTS_Ando, BTS_Ong, ThinFilm_PRL14}; 
Prototypical Bi-based TIs such as Bi$_2$Se$_3$ and Bi$_2$Te$_3$, which are proven 
to possess the topological surface states, 
are still conductive in the bulk due to off stoichiometry \cite{Bi2Se2Te_PRB10}. 

Recently, SmB$_6$ was theoretically predicted to be a TI 
that has the bulk-insulating property \cite{Dzero, Takimoto}. 
SmB$_6$ has long been known as an exemplary heavy-fermion semiconductor, 
or Kondo insulator (KI) \cite{AeppliFisk, Takabatake, Riseborough}. 
It is a mixed valence compound, in which the occupation number of Sm 4$f$ orbital 
lies between 5 and 6 \cite{Cohen_PRL70, Mizumaki_XAS}. 
On cooling, the magnetic moment due to localized $f$ electrons vanishes 
and an activation-type semi-conductivity sets in below $\sim$50 K, 
which are the characteristics of KI governed by strong electron correlations 
\cite{Freericks98, PRL00_PAM}. 
The resistivity attains more than 20 $\Omega$\,cm, 
but unlike in an insulator, it levels off below 4 K, which has been a mystery since the 1960's 
\cite{Menth_1969, Kasuya, PRB78_LowT_JAllen, Cooley_95PRL, Anderson}.  
Transport studies after the theoretical proposal suggest that the low-temperature residual conductivity 
can be the manifestation of topologically metallic surface 
\cite{LowTSurfaceConduction2013, NonlocalTransport2013, NMat14_ThickDep}. 
Spectroscopic studies have reported residual surface states within the Kondo hybridization gap, 
and the compatibility to the topological KI picture is discussed 
\cite{KimuraARPES, Damascelli, Neupane, Feng, Ding_PRB13, 
ARPES_PRX, PointContact_PRX13, SARPES_Ding, PNAS_STM}. 

Regarding that SmB$_6$ possesses the bulk-insulating property and novel metallic surface 
that could be of topological nature, 
it is interesting to search for the response of SmB$_6$ as a bulk insulator 
that can functionalize the novel metallic surface. 
Here, using photoemission spectroscopy implemented 
by a pump-and-probe method, we investigate the response of 
SmB$_6$ to optical pulses. We find that surface photovoltage (SPV) emerges below 
$\sim$90 K accompanied by the evolution of the hybridization gap. 
SPV occurs because of the optically-active surface band bending region
that develops on the edge of semiconductors \cite{SPV_Rev, Kamada, Marsi}. 
The SPV duration exceeds 200 $\mu$s, which is 
good news from opto-electronic application points of 
view, because it is long enough to be detected by electronic devices. 
We also report on the pump-induced variation of the SPV and the electronic 
recovery dynamics, both of which exceeding 100 ps. 

\subsection*{Results}

\begin{figure}[htb]
\begin{center}
\includegraphics[width=7.9cm]{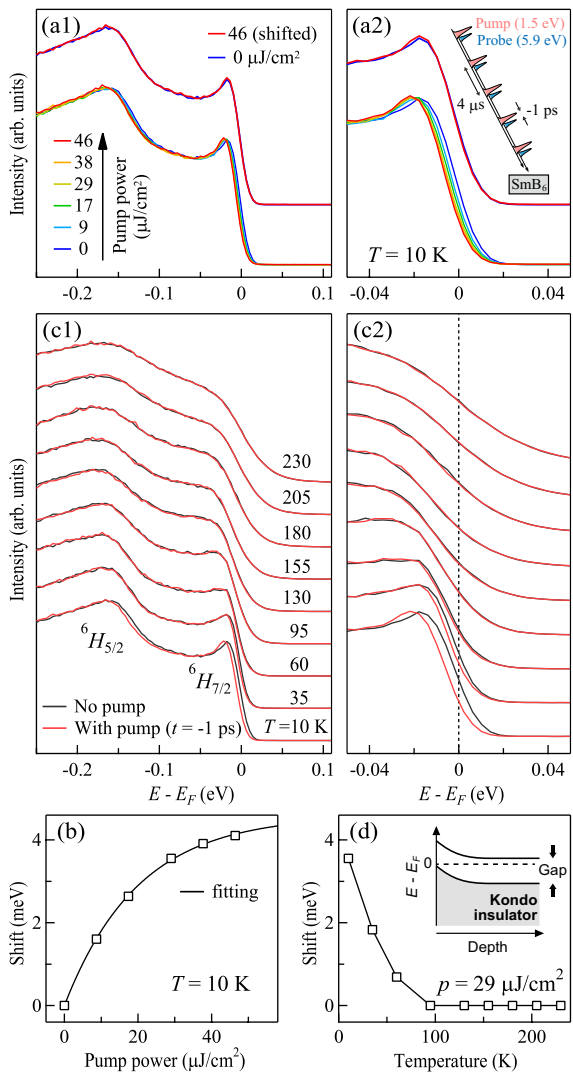}
\caption{\label{fig1} 
{\bf Surface photovoltage effect.}  
(a1, a2) Pump-power dependence of the photovoltaic shift of the spectra  recorded at $T$\,=\,10 K. 
Spectra recorded at $p$\,=\,0 and 46 $\mu$J/cm$^2$ are also displayed with the latter shifted in energy to overlap the former. 
Inset to (a2) schematically shows the pump and probe pulses 
arriving repetitively on the sample with the pump-probe delay of -1 ps. 
(b) Pump-power dependence of the photovoltaic shift at $T$\,=\,10 K. 
(c1, c2) Temperature dependence of the photovoltaic shift. Spectra recorded with and without pump at various temperatures are displayed. Pump power was $p$\,=\,29 $\mu$J/cm$^2$, and the pump-probe delay was set to $t$\,=\,-1 ps.  
(d) Temperature dependence of the shift induced by 
$p$\,=\,29 $\mu$J/cm$^2$. Schematic of the upward surface band bending is also shown. 
Error bars in (b) and (d) are smaller than the marker size.}
\end{center}
\end{figure}

Figure \ref{fig1}(a) shows valence-band spectra of SmB$_6$ recorded at 10 K. 
Here, the sample was driven into a periodic steady state by repetitively irradiating  
the 170-fs pump pulses of 1.5 eV at the interval of 4 $\mu$s (250-kHz repetition); 
see the schematic in Fig.\ \ref{fig1}(a). 
The pump-probe delay was set to $t$\,=\,-1 ps, so that the probing by the 5.9 eV pulse was done 
just before the arrival of the pump, or equivalently, 4 $\mu$s after the arrival of the previous pump. 
As the pump power $p$ was increased, the spectrum consisting of 
$H_{7/2}$ and $H_{5/2}$ peaks located at -20 and -150 meV, respectively, 
shifted into the occupied side (lower energies) with negligible modification in the spectral shape; see Fig.\ \ref{fig1}(a), in which the spectrum recorded at $p$ = 46 $\mu$J/cm$^2$ nicely overlaps to that recorded without pump after some shift in energy. The pump-induced shift  $\delta$ 
as a function of $p$ is plotted in Fig.\ \ref{fig1}(b): For the derivation of $\delta$, see Methods. The  $\delta$-$p$ curve shows 
the tendency of saturation with increasing $p$. Through fitting by a function 
$\delta = \delta_0(1 - e^{-p/p_0})$, the saturation value $\delta_0$ at 10 K 
is estimated to be 4.6 meV ($p_0$ is a fitting parameter). 
We did not observe the broadening of the spectrum in the pump-power range 
investigated herein, indicating that the heating of the sample due to the irradiation of the  
pump beam was negligibly small.

The pump-induced shift of the spectrum is attributed to the SPV effect \cite{SPV_Rev, Kamada, Marsi}, which is similar to the photovoltaic effect occurring in the interfacial band-bending region 
in solar cells and photo-diodes. 
SPV usually occurs because of the photo-induced relaxation of the surface band bending. 
Therefore, the pump-induced shift into the occupied side indicates that the bulk is slightly $n$-type doped to exhibit an upward surface band bending on the edge, 
and that this bending relaxes under the pump-beam irradiation; 
see the schematic in Fig.\ \ref{fig1}(d). 
Concerning that the SPV is at most the size of the band bending, 
it is reasonable that the saturation value $\delta_0$ is comparable to the 
activation-gap size of 3 meV and is smaller than the hybridization gap of 15-20 meV 
reported to date \cite{R99_3_19meV}. 
We can exclude the possibility of photo-induced increase of the band bending, or the photovoltage inversion, 
because it can occur only when the photon energy of the irradiation is less than the band gap \cite{SPV_Rev_Gatos}.  
We can also exclude the charging effect as the origin of the pump-induced shift, because the the pump
generated negligibly small amount of multi-photon photoelectrons compared to the amount of
photoelectrons generated by the probe. Therefore, if the charging effect existed, it should have been
already prominent just by the probe-beam irradiation. 
Moreover, the charging-induced shift do not have reasons 
to exhibit saturation behavior as seen in the $\delta$-$p$ curve presented in Fig.\ \ref{fig1}(b). 
Signatures of possible pump-induced variations in the surface states were not observed in the spectrum, which would have manifested as changes additional to the photovoltaic shift of the states within the surface band bending region.

Next, we show the temperature dependence of the photovoltaic shift. In Fig.\ \ref{fig1}(c), we show spectra recorded with and without pump at various temperatures. Here, $p$ and $t$ were fixed to 29 $\mu$J/cm$^2$ and -1 ps, respectively. As the temperature was increased, the photovoltaic shift diminished and became negligible above $\sim$90 K ($\equiv T^*$). $T^*$ nicely coincides to the temperature around which the hybridization gap diminishes \cite{PointContact_PRX13}. 
This shows that the photovoltaic effect emerges upon the hybridization-gap opening 
and coupled evolution of the surface band bending. 
$\delta$ as a function of temperature is plotted in Fig.\ \ref{fig1}(d). 
The $\delta$-$T$ curve is concave 
($\partial^2\delta/\partial T^2 > 0$), and is contrasted to 
the gap evolution in a second-order transition. 
This may reflect the crossover nature of the gap opening in 
the periodic Anderson model, in which the temperature-dependent 
renormalization follows $\propto$\,$-$$\log T$ behavior \cite{PRL00_PAM, CeIrIn5}.

\begin{figure}[htb]
\begin{center}
\includegraphics[width=8.2cm]{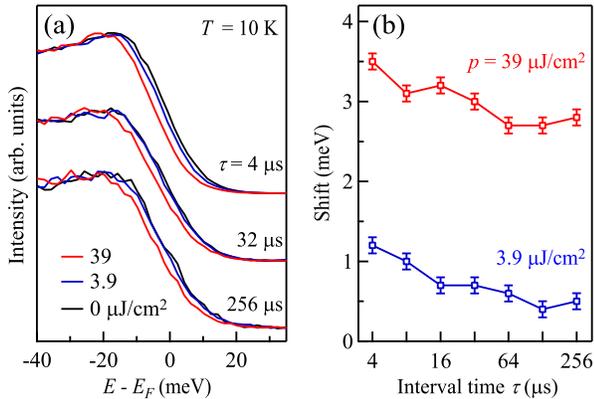}
\caption{\label{fig2} 
{\bf Duration of the surface photovoltage effect. }  
(a) Spectra recorded under the repetitive pump pulses arriving at the 
intervals of $\tau =$ 4, 32, and 256 $\mu$s. Spectra were recorded at $T$ = 10 K and $t$\,=\,-1 ps. 
(b) Photovoltaic shift $\delta$ as functions of the interval time $\tau$. The error bars represent typical standard deviations. }
\end{center}
\end{figure}

The photovoltaic shift observed at $t$ = -1 ps, 
or 4 $\mu$s after the previous pulse, 
indicates that the duration of the photovoltage is 
exceeding 4 $\mu$s. 
In order to obtain in-depth information, we investigated the repetition-rate 
(interval-time $\tau$) dependence of the pump-induced shift. 
To this end, we repetitively deflected out the pulse out of the 
250 kHz laser by using a pulse picker and used the deflected pulse in 
the pump-and-probe measurement. 
In this way, we can set the interval time $\tau$ to 
4 $\mu$s $\times$ 2$^n$  ($n$ = 0, 1, ..., 6)  without changing 
the intensity per pulse. 
As shown in Figs.\ \ref{fig2}(a) and \ref{fig2}(b), the pump-induced shift monotonically 
decreased as $\tau$ was increased. 
This is in agreement with the picture that the remaining SPV responses to 
the preceding pulses became small as the interval of 
the pulses was increased. 
However, $\delta$ did not decrease exponentially with $\tau$, 
and the SPV duration exceeded 200 $\mu$s. 
This shows that 
the SPV response induced by the intense laser pulse 
is in the saturation regime, so that $\delta$ is not 
a simple addition of the individual responses to each pulse: 
Note, the SPV is limited by the amount of the surface band bending built in at equilibrium. 
The saturation behavior is also 
noticed by the fact that the SPV is still sizable even when the average 
pump power ($p$ multiplied by the repetition rate) is reduced to 1/64 [Fig.\ \ref{fig2}(b)].

\begin{figure*}[htb]
\begin{center}
\includegraphics[width=17cm]{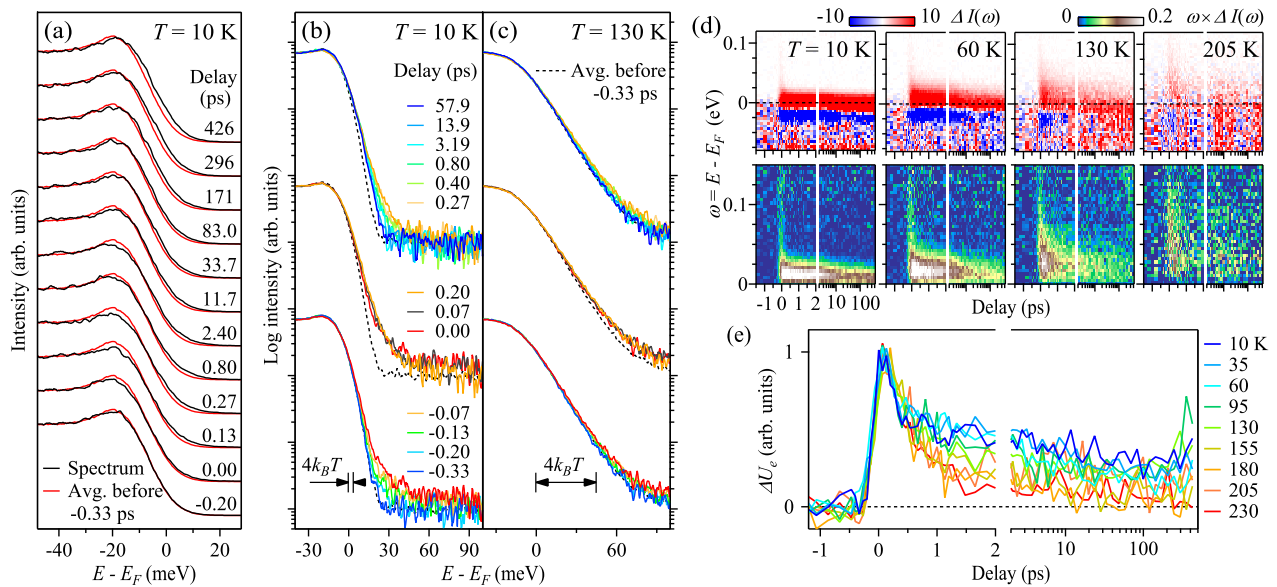}
\caption{\label{fig3} 
{\bf Initial dynamics induced by the pump pulse. } 
(a) Spectra recorded at $T$ = 10 K, $p$ = 29 $\mu$J/cm$^2$, and $\tau$ = 4 $\mu$s at several delays. 
The averaged spectrum before -0.33 ps is also overlaid on each spectrum. The shift of the spectra into the unoccupied side becomes prominent at $t \gtrsim$ 100 ps, 
which is attributed to the variation of the surface photovoltage induced by the pump pulse.     
(b, c) Spectra in a logarithmic scale plot at $T$ = 10 and 130 K, respectively. 
The energy scale of 4$k_BT$ is also displayed. 
(d) Pump-induced difference of the spectrum (upper panels) and excess-energy distributions 
(lower panels) mapped in the 
$\omega$\,-\,$t$ plane ($\omega \equiv E - E_F$) at various temperatures. 
(e) Dissipation of the excess-electronic energy (see, text) at various temperatures.} 
\end{center}
\end{figure*}

Next, we investigate the initial dynamics induced by the pump pulse 
in the femto-to-picoseconds. 
Here, we find 
ultrafast electron redistribution on arrival of the pump, 
subsequent recovery of the electronic system that persists for $>$100 ps, and 
gradual change of the photovoltage after $\sim$100 ps, as we describe below. 
 
Figures \ref{fig3}(a) and \ref{fig3}(b) show photoemission spectra at various pump-probe delays 
displayed in linear- and logarithmic-scale plots, respectively. Here, $T$, $p$, and $\tau$ were 
10 K, 29$\mu$J/cm$^2$, and 4 $\mu$s, respectively. 
On arrival of the pump (-0.3\,$<$\,$t$\,$<$\,0 ps), the spectral intensity 
increases in the unoccupied side. Seen in the logarithmic-scale 
plot [Fig.\ \ref{fig3}(b)], we discern a plateau feature in the unoccupied side, 
which is similar to those observed in the time-resolved photoemission spectra of 
Au \cite{Au} and graphite \cite{Ishida_HOPG}. 
This indicates that a non-thermal electron distribution (non-Fermi-Dirac distribution) is realized 
on arrival of the pump. 
Then, the plateau feature turns over into an exponential tailing at 
0\,$<$\,$t$\,$<$ 0.2 ps [Fig.\ \ref{fig3}(b)]. This indicates that the 
electronic distribution turned into Fermi-Dirac distribution function, and the electronic 
temperature became definable. 
After 0.2 ps, the slope of the exponential tailing gradually 
recovers to the value before the pump with the time scale 
exceeding 100 ps. This indicates that the cooling of the 
electronic system occurred on the $>$100-ps time scale. 
The exponential tailing is not exceeding the energy scale of $k_BT^*$\,$\sim$\,30 meV ($k_B$ is the Boltzmann constant), 
so that the electronic temperature is estimated to be less than $T^*$ at $>$0.2 ps. This indicates that the pump did not collapse the hybridization gap or that the gap had already recovered by 0.2 ps. 

In Fig.\ \ref{fig3}(a), we also observe that the spectra are gradually shifted into the 
unoccupied side at $\gtrsim$100 ps. 
This is attributed to the pump-induced variation of the photovoltage in the periodic steady state. 
SPV varies after the photo-excited electrons and holes drift into opposite 
directions in the surface-band-bending region, and therefore, a delay can exist 
in the SPV response \cite{SPV_Rev, Kamada, Marsi}. 
The shift of the spectra into the unoccupied side indicates that the 
amount of band bending is increasing at $\gtrsim$100 ps.
In the SPV dynamics in semiconductors \cite{Kamada, Marsi}, 
there are fast (sub nano-second) and slow (over 10 nano-second) 
components in the SPV recovery. 
The fast component is known to be sensitive to surface conditions and 
surface states \cite{Marsi}. It is also reported that when photo-induced 
depopulation of surface states occur, photovoltage can increase 
the amount of band bending rather than decreasing it 
\cite{SPV_Rev_Gatos, SPV_Rev, Marsi}.  
Therefore, the pump-induced variation of SPV 
at $>$100 ps may be involving interesting dynamics in the surface states of SmB$_6$, 
although the detailed mechanism is not clear at present.

At $T$ = 130 K [Fig.\ \ref{fig3}(c)], the 
pump-induced shift at $>$100 ps was hardly detected, 
while the cooling of the electronic system 
was still taking place for $>$100 ps (see later). 
The absence of the shift at $>$100 ps is consistent 
with the fact that the pump-induced photovoltaic effect occurs only 
below the characteristic temperature $T^*$ as shown in Fig.\ \ref{fig1}. 
The pump-induced change in the spectrum 
at $>$0.2 ps occurs in a wider energy range 
set by the thermal broadening of the spectra, 
see the 4$k_BT$ energy scale 
displayed in 
Fig.\ \ref{fig3}(c). The pump-induced changes 
occurring farther into the unoccupied side at higher temperatures 
can also be seen in Fig.\ \ref{fig3}(d), in which the  
intensity maps of the pump-induced change 
$\varDelta I(\omega, t) = I(\omega, t) - I_0(\omega)$ 
and excess electronic energy distribution $\omega\varDelta I(\omega, t)$ 
at various temperatures are displayed in the upper and lower panels, respectively 
[$I_0(\omega)$ is the average of the 
spectrum recorded at -3\,$<$\,$t$\,$<$\,-0.33 ps]. 
For the full set of pump-and-probe photoemission spectra 
taken at various temperatures, 
see a supplementary movie file 
in which the frames of  the spectra in linear and logarithmic scale, and that of 
the difference spectra are shown. 

\subsection*{Discussion}

In general, the recovery dynamics after the pump is considered to be 
very different between metals and semiconductors. 
In metals, the recovery time of the electronic system is typically of 
1 ps \cite{Allen, Brorson, Au}, while strong electron correlations may 
delay the recovery to some extent \cite{WernerPRB12}. 
When the electronic dynamics is coupled to the lattice degrees of freedom, 
the recovery can be bottlenecked around $\sim$1 ps, because the 
slow heat dissipation from the lattice may 
govern the recovery thereafter \cite{Perfetti, Ishida_HOPG}. 
In gapped systems such as 
semiconductors and superconductors,  
the recovery time can exceed 100 ps 
because the electron-hole annihilation 
across the gap emits some boson modes that re-excites  
electron-hole pairs \cite{RT, Haight, MgB2, PRB12_Ono}; in effect, the recovery is slowed. 

Having outlined above, we look into the indications of the slow 
electronic recovery as evidenced by the slope of the spectral tailing not fully recovered even at 100 ps. 
First, the long recovery time observed at $T<T^*$ indicates that, 
even though there are in-gap states, the 
photo-excited electron-hole pairs are protected from 
fast recombinations of $\sim$1 ps seen in metals. 
This is in accord with the picture that 
the in-gap states are localized on surface, and that they are spatially separated 
from the electron-hole pairs in the bulk. 
Next, the recovery persisting for $>$100 ps even at $T > T^*$ may be 
reflecting the crossover nature of the 
hybridization gap, so that the semiconductor characteristics 
in the recovery may still remain above $T^*$. Alternatively, 
the lattice degrees of freedom may be slowing the recovery. 
In order to see whether a bottleneck exists in the recovery, we plot 
$\varDelta U_e(t)\equiv \int_{\omega > 0} \omega\varDelta I(\omega, t) d\omega$ 
at various temperatures in Fig.\ \ref{fig3}(e). 
$\varDelta U_e(t)$ is a good measure of how the excess electronic energy dissipates with time 
(This is a good measure even at $T < T^*$ if $t\lesssim$100 ps, 
because in that time region, the photovoltaic shift is still small).  
Clearly, there is a crossover in the  
electronic recovery at $\sim$0.5 ps. 
This is the bottleneck effect, which is usually attributed to the thermalization between electronic and coupled phonon systems in a  two-temperature model scheme \cite{Allen, Perfetti}, 
although the microscopic validation of the model as well as the mechanism of the bottleneck are still under discussion \cite{Ishida_HOPG, PRB14_Kabanov}. The relaxation after $\sim$0.5 ps can be reflecting the slow heat transfer from the slightly hot electronic system to the slightly cool lattice \cite{PRB14_Kabanov}. 
Whatever the origin may be, the bottleneck at $\sim$0.5 ps observed also at $T\,>\,T^*$ shows that the slowing of the recovery is not unique to temperatures below $T^*$.

The electronic recovery time exceeding 100 ps was also observed in a 
pump-and-probe photoemission study of 
Bi-based TI that has bulk resistivity as high as $\sim$4 $\Omega$\,cm at 4 K \cite{MarsiTI}.  
In contrast to the present study, however, SPV effect was not reported therein. 
Our study thus demonstrates the co-existence of the optically active 
band bending region and novel metallic state on the edge of the highly bulk resistive SmB$_6$. 
The SPV effect can be utilized as an opto-electronic device function 
such as optical gating of the novel metallic surface for extracting the spin currents, 
which would be readily detected by means of electronic devices 
concerning its long duration exceeding 200 $\mu$s.  

\subsection*{Acknowledgement}
The authors acknowledge T.~Nakamura and M.~Endo for technical help, and  
S.-Y.~Xu, M.~Neupane, M.~Z. Hasan, T.~Oka, and P.~Werner for discussion. 
This work was supported by JSPS through FIRST program, by Photon and 
Quantum Basic Research Coordinated Development Program from MEXT, 
and by JSPS KAKENHI (20102004, 23540413, and 26800165).

\subsection*{Author contributions}

Y.I.\ and T.O.\ performed the experiments in assist of T.S.\ and M.O.; 
T.O.\ and Y.I.\ made the repetition-rate of the laser system changeable 
under the direction by Y.K.; F.I.\ grew high-quality single crystals 
under the supervision of T.T.; Y.I.\ analyzed the data and wrote the manuscript; 
S.S.\ supervised the project; All authors discussed the results 
and commented on the manuscript.

\subsection*{Competing financial interests}

The authors declare no competing financial interests.

\subsection*{Methods}

Single crystalline samples of SmB$_6$ were grown by floating-zoned method 
using a 4-xenon-lamp image furnace, which was also used for single 
crystal growth of KI YbB$_{12}$ \cite{Iga}.
Samples of 2\,$\times$\,2$\times$\,6\,mm$^3$ in size cut out along [001] were cleaved in the spectrometer at the base pressure of 5$\times$10$^{-11}$ Torr. Time-resolved photoemission spectroscopy was done in a pump-and-probe configuration \cite{Ishida_RSI}. 
The laser pulse (1.5 eV and 170 fs duration) delivered from a Ti:Sapphire laser system 
operating at 250 kHz repetition (Coherent RegA 9000) was split into two: 
One pulse is used as a pump, and the other was up-converted into 5.9 eV 
and used as a photoemission source (probe). The repetition rate of the laser pulses 
was discretely changeable by using 
a pulse picker consisting of an acoustic optical modulator. 
The pump-and-probe delay $t$ was controlled by changing 
the optical path length of the pump beam line. 
The delay origin $t$ = 0 and time resolution (300 fs) were determined by recording the pump-probe photoemission signal of graphite attached next to the sample \cite{Ishida_HOPG}. 
The diameters of the pump and probe beams on the sample position were 0.5 and 0.3 mm, respectively. 
The Fermi energy $E_F$ and the energy resolution (18 meV) were determined 
by recording the Fermi cutoff of gold in electrical contact to the sample and 
the VG Scienta R4000 analyzer. 

The datasets of the spectra presented in Figs.\ \ref{fig1}(a), \ref{fig1}(c), \ref{fig2}(a) and \ref{fig3}(a)\,-\,\ref{fig3}(c) were acquired via automated sequences, in which parameters such as $p$ and $t$ were repetitively cycled over the specified values during the acquisition; see Ref.\ \cite{Ishida_RSI}. For example, the set of spectra presented in Fig.\ \ref{fig1}(a) were accumulated while cycling $p$ from 0 to 46 $\mu$J/cm$^2$; for those presented in Fig.\ \ref{fig2}(a), $p$ was cycled as 0 $\to$ 3.9 $\to$ 39 $\mu$J/cm$^2$ under the respective repetition rates. In this way, we can accumulate the spectra until sufficient signal-to-noise ratio ($S/N$) is simultaneously achieved in the whole dataset. Such a dataset acquisition is advantageous because each of the spectrum in the raw dataset is normalized to the acquisition time and has the same $S/N$. This allowed us a normalization-free dataset analyses when deriving various quantities such as $\delta$ and $\varDelta U_e$. The spectral shift $\delta$ [Figs.\ \ref{fig1}(b) and \ref{fig2}(b)] was determined such that $\delta$ minimized $F(\delta) = \int |I(\omega-\delta) - \tilde{I}(\omega)|^2\,d\omega$, where $I(\omega)$ and $\tilde{I}(\omega)$ are the raw spectra recorded with and without pump, respectively.

\end{document}